\documentclass[12pt]{iopart}
\begin{document}
\title[Geometric frustration in the Ising-Heisenberg diamond chains]
{Geometric frustration in the class of exactly solvable Ising-Heisenberg diamond chains}
\author{Lucia \v{C}anov\'a, Jozef Stre\v{c}ka\footnote{Corresponding author: jozkos@pobox.sk} 
and Michal Ja\v{s}\v{c}ur}
\address{Department of Theoretical Physics and Astrophysics, Faculty of Science, \\
P. J. \v{S}af\'arik University, Park Angelinum 9, 040 01 Ko\v{s}ice, Slovak Republic}
\ead{lucia.canova@post.sk}
\begin{abstract}
Ground-state and finite-temperature properties of the mixed spin-$\frac12$ and spin-$S$ Ising-Heisenberg diamond chains are examined within an exact analytical approach based on the generalized decoration-iteration map. A particular emphasis is laid on the investigation of the effect of geometric frustration, which is generated by the competition between Heisenberg- and Ising-type exchange interactions. It is found that an interplay between the geometric frustration and quantum effects gives rise to several quantum ground states with entangled spin states in addition to some semi-classically ordered ones. Among the most interesting results to emerge from our study one could mention a rigorous evidence for quantized plateux in magnetization curves, an appearance of the round minimum in the thermal dependence of susceptibility times temperature data, double-peak zero-field specific heat curves, or an enhanced magnetocaloric effect when the frustration comes into play. The triple-peak specific heat curve is also detected when applying small external field to the system driven by the frustration into the disordered state.
\end{abstract}
\begin{center}
\submitto{\JPCM}
\end{center}

\section{Introduction}

Over the last three decades, the low-dimensional quantum spin models with competing (frustrated) interactions 
have attracted considerable research interest especially due to their extraordinary diverse ground-state behaviour. \textit{Geometrically frustrated spin systems} constitute a special sub-class of the frustrated models that can be distinguished by incapability of spins, inherent in their lattice positions, to simultaneously minimize the ground-state energy of each individual spin-spin interaction \cite{Tou77}. 
As a rule, the quantum spin systems affected by a rather strong geometric frustration often exhibit 
an exotic non-magnetic ground state (which does not have its classical analogue) in addition to a rich 
variety of the semi-classically ordered ones \cite{Gre01}. 
It is worthy to notice, moreover, that a subtle interplay between the geometric frustration and quantum fluctuations gives rise to a number of intriguing phenomena including the macroscopic degeneracy of ground state \cite{Die04}, order by disorder effect \cite{Vil80}, chirality \cite{Vil77}, quantum phase transitions \cite{Sach99}-\cite{Rich04}, quantized plateaux in the magnetization curves \cite{Rich04}-\cite{Hon04}, double-peak specific heat curves \cite{Kub93}-\cite{Mai98}, enhanced magnetocaloric effect \cite{Zhi03}-\cite{Hon05}, etc. 

Despite a significant amount of effort, there are only few frustrated spin-$\frac12$ quantum Heisenberg models,
such as Majumdar-Ghosh model on double chain \cite{Maj69a}-\cite{Bro80}, sawtooth ($\Delta$) chain \cite{Dou89}-\cite{Sen96}, or Shastry-Sutherland model \cite{Sha81}, for which precise analytic 
solution is available leastwise for the ground state. Nevertheless, it should be pointed out 
that frustrated quantum systems are in general rather difficult to deal with, since extensive numerical methods must be used in order to obtain a reliable estimate of their magnetic properties. From this point 
of view, the one-dimensional (1D) frustrated spin systems are the simplest systems with respect to accurate treatment. Of these systems, the spin-$\frac12$ quantum Heisenberg model with \textit{diamond chain} 
topology is currently actively engaged in the investigation of geometric frustration. Interestingly, 
this simple quantum system turned out to have rather complex ground state; apart from 
the usual ferrimagnetic phase there are in fact several quantum dimerized and plaquette states involved 
in the zero-field ground-state phase diagram  \cite{Tak96}-\cite{Wan02}. Further studies 
devoted to the spin-$\frac12$ quantum Heisenberg model on the diamond chain have provided fairly accurate 
results for the ground-state phase diagram in a presence of the external field \cite{Ton00}-\cite{Oka03}, 
the spin gap \cite{San00}, the magnetization and susceptibility \cite{Hon01}. Another remarkable 
finding relates to the observation of an inversion phenomenon, which can be induced in the 
frustrated diamond chain through the exchange anisotropy \cite{Oka02, Oka05}. Note that the ground 
state and thermodynamics of the mixed-spin diamond chains containing also higher-spin 
sites have already been particularly examined as well \cite{Nig97}-\cite{Vit02b}.

It is worthwhile to remark that 1D frustrated spin systems have initially been introduced purely 
as toy models suitable for investigating the effect of spin frustration. However, a recent progress 
achieved in the design and controlled synthesis of molecular-based magnetic materials afforded another 
stimulus for testing 1D frustrated spin systems by overcoming the lack of desirable model compounds. 
As a matter of fact, the rapidly expanding field of molecular engineering have led to fabrication
of several coordination polymers, which can be regarded as genuive examples of the frustrated 
spin models \cite{Ram94}. With the help of structural data known long ago \cite{Zig72}, 
Kikuchi and co-workers \cite{Kik03}-\cite{Oku05} 
have recognized Cu$_3$(CO$_3$)$_2$(OH)$_2$ (azurite) as an appropriate candidate for the diamond chain compound. Experimental data measured for the high-field magnetization, susceptibility, specific heat, 
NMR \cite{Kik03}-\cite{Kik05} and ESR data \cite{Oht03}-\cite{Oku05} have indeed confirmed Kikuchi's conjecture and it is now quite well established that the azurite represents the actual material for the frustrated diamond chain. It should be also mentioned that sufficiently strong frustration found in the azurite clearly manifests itself in its quantum features; the high-field magnetization shows quantum plateau at one third of the saturation magnetization, the susceptibility turned out to have two round peaks at 
relatively low temperatures and ESR data have proven the gap-less excitation spectrum.
Theoretical interest focused on the diamond chain structure also enhances its another 
experimental realizations provided by the polymeric compounds such as Cu$_2$OSO$_4$ \cite{Bel91} 
and M$_3$(OH)$_2$ (M = Ni, Co, Mn) \cite{Gui02}-\cite{Ton05}.

With this background, we shall investigate in the present article a simplified version of the frustrated
Ising-Heisenberg diamond chain, which can exactly be solved by applying an accurate map based on the 
generalized decoration-iteration transformation \cite{Fis59, Syo72}. It is noteworthy that this 
relatively simple and straightforward analytical approach has recently been adapted to study 
an appearance of quantized plateaux in the magnetization process of the trimerized  \cite{Str02, Str03} 
and tetramerized \cite{Str04, Str05} Ising-Heisenberg linear chains. As we shall show hereafter, the same strategy can also be used to explore the frustrated Ising-Heisenberg diamond chain, which represents 
another particular realization of models tractable within the generalized algebraic maps. The main goal 
of present work is to exploit a grand advance of the method used for obtaining exact results for all 
possible ground states as well as all basic thermodynamic quantities. 

The rest of this paper is organized as follows. The model Hamiltonian and major features of the mapping 
method are presented in Section 2. Section 3 is divided into two sub-sections. In the former one, we 
provide rigorous results for the spin-$\frac12$ Ising-Heisenberg diamond chain, while the latter one comprises 
magnetic data of the mixed spin-$\frac12$ and spin-1 Ising-Heisenberg diamond chain. Altogether, 
exact results for ground-state phase diagrams, magnetization, entropy, susceptibility and specific heat
are derived and particularly discussed for both investigated diamond chains. A cooling rate of adiabatic demagnetization is also explored in connection with the enhanced magnetocaloric effect. Finally, some concluding remarks are drawn in Section 4.

\section{Model and its exact solution}

Let us begin by considering two kinds of spins regularly distributed on the 1D lattice composed 
of the diamond-shaped units as diagrammatically depicted in Fig.~1. To ensure an exact 
tractability of this spin system, we shall further suppose that each diamond-shaped plaquette consists 
of two dumbbell \textit{Heisenberg} spins ($S$), which are placed in between two \textit{Ising} spins 
($\mu$) residing corner-sharing positions on the diamond motifs. For further convenience, 
we shall write the total Hamiltonian as a sum over plaquette Hamiltonians, i.e. $\hat{{\cal{H}}} = \sum_{k = 1}^N \hat{{\cal{H}}}_k$, where each plaquette Hamiltonian $\hat{{\cal{H}}}_k$ involves all interaction terms associated with one diamond-shaped unit (see Fig. 1)
\begin{eqnarray}
\label{eq:Hk}
\fl \hat{{\cal{H}}}_k &= J_{\rm H} [\Delta(\hat{S}_{3k-1}^{x} \hat{S}_{3k}^{x} + \hat{S}_{3k-1}^{y} \hat{S}_{3k}^{y}) + \hat{S}_{3k-1}^{z}\hat{S}_{3k}^{z}] + J_{\rm I} (\hat{S}_{3k-1}^{z} + \hat{S}_{3k}^{z})(\hat{\mu}_{3k-2}^{z} + \hat{\mu}_{3k+1}^{z}) \nonumber \\
\fl &- H_{\rm H} (\hat{S}_{3k-1}^{z} + \hat{S}_{3k}^{z}) 
 - H_{\rm I} (\hat{\mu}_{3k-2}^{z} + \hat{\mu}_{3k+1}^{z})/2.
\end{eqnarray} 
Above, $\hat{\mu}_{k}^{z}$ and $ \hat{S}_{k}^{\alpha}$ ($\alpha = x, y, z$) denote spatial components 
of the spin-$\frac12$ and spin-$S$ operators, the parameter $J_{\rm H}$ stands for the anisotropic XXZ 
interaction between the nearest-neighbouring Heisenberg (dumbbell) spins and the parameter $J_{\rm I}$ accounts for the Ising-type coupling between the Heisenberg spins and their nearest Ising neighbours. 
The parameter $\Delta$ allows to control the Heisenberg exchange interaction $J_{\rm H}$ between the easy-axis ($\Delta<1$) and easy-plane ($\Delta>1$) regime and finally, the last two terms incorporate coupling of 
the Ising and Heisenberg spins to an external longitudinal magnetic field $H_{\rm I}$ and $H_{\rm H}$, respectively. 
 
The crucial point of our calculation represents calculation of the partition function. By making use 
of the commutation rule between different plaquette Hamiltonians, i.e. by exploiting $[\hat{{\cal{H}}}_k, \hat{{\cal{H}}}_l] = 0$ valid for each $k \neq l$, the partition function of Ising-Heisenberg diamond chain 
can be partially factorized into the following product
\begin{eqnarray}
\label{eq:Z}
{\cal{Z}} &=& \sum_{\{\mu_i \}}\prod_{k = 1}^{N} \mathrm{Tr}_k \exp(-\beta \hat{{\cal{H}}}_{k}),
\end{eqnarray}
where $\beta = 1/(k_{\rm B}T)$, $k_{\rm B}$ is being Boltzmann's constant, $T$ the absolute temperature, 
the symbol $\sum_{\{\mu_i \}}$ means a summation over all available configurations of Ising spins $\{ \mu_i \}$
and $\mathrm{Tr}_k$ stands for a trace over spin degrees of freedom of two dumbbell Heisenberg spins 
belonging to $k$th diamond unit. After performing this partial trace, the structure of relation 
(\ref{eq:Z}) immediately implies a possibility of applying the generalized decoration-iteration 
mapping transformation \cite{Fis59, Syo72} 
\begin{eqnarray}
\label{eq:DIT}
\mathrm{Tr}_k \exp(- \beta\hat{{\cal{H}}}_{k}) &=& 
\exp \bigl[ \beta H_{\rm I} (\mu_{3k-2}^{z} + \mu_{3k+1}^{z})/2 \bigr ] 
G \bigl[ \beta J_{\rm I} (\mu_{3k-2}^{z} + \mu_{3k+1}^{z})/2 \bigr ] \nonumber \\
&=& A \exp \bigl[\beta R \mu_{3k-2}^{z} \mu_{3k+1}^{z} 
    + \beta H_{0}(\mu_{3k-2}^{z} + \mu_{3k+1}^{z})/2 \bigr],
\end{eqnarray}
which converts the Ising-Heisenberg diamond chain into the uniform spin-$\frac12$ Ising linear chain with 
an effective nearest-neighbour coupling $R$ and an effective external field $H_0$. The expression $G(x)$ 
in the first line of Eq. (\ref{eq:DIT}) depends on the spin of Heisenberg atoms and its explicit form 
is given in Appendix for two particular spin cases $S = \frac12$ and $1$. It is noteworthy that a general validity of the mapping relation (\ref{eq:DIT}) necessitates a self-consistency condition to be satisfied, what means, that it must hold independently of spin states of both Ising spins $\mu_{3k-2}$ and $\mu_{3k+1}$. 
It can readily be proved that a substitution of four possible configurations of the Ising spins $\mu_{3k-2}$ and $\mu_{3k+1}$ into the formula (\ref{eq:DIT}) indeed gives just three independent equations, 
which unambiguously determine the unknown mapping parameters $A$, $R$ and $H_0$ 
\begin{eqnarray}  
\label{eq:AR}
A = \bigl(G_1 G_2 G_3^2 \bigr)^{1/4}, 
\quad
\beta R = \ln \Biggl( \frac{G_1 G_2}{G_3^2} \Biggr),
\quad
\beta H_0 = \beta H_{\rm I} + \ln \Biggl( \frac{G_1}{G_2} \Biggr).
\end{eqnarray}
Here, we have defined the functions $G_1 = G(\beta J_{\rm I})$, $G_2 = G(-\beta J_{\rm I})$ and 
$G_3 = G(0)$ in order to write the transformation parameters $A$, $R$ and $H_0$ in more abbreviated 
and general form. Now, a direct substitution of the transformation (\ref{eq:DIT}) into the expression (\ref{eq:Z}) yields the equality
\begin{equation} 
\label{eq:ZdZ0}
{\cal Z} (\beta, J_{\rm I} , J_{\rm H}, \Delta, H_{\rm I}, H_{\rm H}) = A^{N} {\cal Z}_{0}(\beta, R, H_0),
\end{equation}
which establishes an exact mapping relationship between the partition function $\cal{Z}$ of the Ising-Heisenberg diamond chain and, respectively, the partition function ${\cal Z}_{0}$ of the uniform spin-$\frac12$ Ising linear chain defined by means of the nearest-neighbour coupling $R$ and the effective field $H_0$. Notice that the exact solution for the partition function of the latter system is well-known 
(see for instance Ref. \cite{Lav99}) and hence, the relation (\ref{eq:ZdZ0}) can readily be utilized for calculation of some important quantities (magnetization, entropy, specific heat, susceptibility) by making use of the standard thermodynamical-statistical relations. It should be mentioned, however, 
that there also exists another alternative approach that is of particular importance if some relevant 
physical quantity cannot be obtained within this procedure. Actually, the problem connected with the calculation of correlation functions and/or quadrupolar moment can be simply avoided by employing the following exact spin identities  
\begin{eqnarray}
\fl \langle f_{1} (\hat{\mu}_1^{z}, ... , \hat{\mu}_{3k-2}^{z}, ... , \hat{\mu}_{3N-2}^{z}) 
\rangle = \langle f_{1} (\hat{\mu}_1^{z}, ... , \hat{\mu}_{3k-2}^{z}, ... , \hat{\mu}_{3N-2}^{z}) \rangle_{0}, \label{eq:f1} \\
\fl \langle f_{2}(\hat{\mu}_{3k-2}^{z}, \hat{S}_{3k-1}^{\alpha}, \hat{S}_{3k}^{\gamma}, \hat{\mu}_{3k+1}^{z})      \rangle =  \Bigg\langle \frac{\Tr_{k} f_{2} (\hat{\mu}_{3k-2}^{z}, \hat{S}_{3k-1}^{\alpha}, \hat{S}_{3k}^{\gamma}, \hat{\mu}_{3k+1}^{z}) \exp(-\beta \hat{{\cal{H}}}_{k})}
{\Tr_k \exp(-\beta \hat{{\mathcal{H}}}_{k})} \Bigg\rangle.  
\label{eq:f2}
\end{eqnarray}
In above, the function $f_{1}$ depends exclusively on the Ising spin variables $\{ \mu_i \}$, while the 
function $f_{2}$ may depend on any spin variable belonging to $k$th diamond plaquette. 
The superscripts $\alpha, \gamma \in (x, y, z)$ label spatial components of the appropriate spin operators 
and finally, the symbols $\langle ... \rangle$ and $\langle ... \rangle_{0}$ stand for the standard canonical 
average performed over the ensemble defined on the Ising-Heisenberg diamond chain and its corresponding 
Ising chain, respectively. By applying the exact spin identities (\ref{eq:f1}) and (\ref{eq:f2}), one 
easily attains rigorous results for the sub-lattice magnetization $m_{\rm i}^z$ and $m_{\rm h}^z$ reduced per one Ising and Heisenberg spin, respectively, the pair correlation functions $c_{\rm hh}^{xx}$, $c_{\rm hh}^{zz}$, $c_{\rm ih}^{zz}$, as well as, the quadrupolar moment $q_{\rm hh}^{zz}$
\begin{eqnarray} 
m_{\rm i}^z &\equiv&                     
\langle \hat{\mu}_{3k-2}^{z} \rangle
= \langle \hat{\mu}_{3k-2}^{z} \rangle_{0} \equiv m_0,  \label{eq:miz} \\
c_{\rm ii}^z &\equiv&                     
\langle \hat{\mu}_{3k-2}^{z} \hat{\mu}_{3k+1}^{z} \rangle
= \langle \hat{\mu}_{3k-2}^{z} \hat{\mu}_{3k+1}^{z} \rangle_0 \equiv \varepsilon_0,  \label{eq:viz} \\
m_{\rm h}^z &\equiv& \langle \hat{S}_{3k}^{z}\rangle
= K_1/4 + m_0 L_1 + \varepsilon_0 M_1,  \label{eq:mhz} \\
c_{\rm hh}^{zz} &\equiv&                     
\langle \hat{S}_{3k-1}^{z} \hat{S}_{3k}^{z} \rangle
= K_2/4 + m_0 L_2 + \varepsilon_0 M_2,  \label{eq:qhz}\\
c_{\rm hh}^{xx} &\equiv&                     
\langle \hat{S}_{3k-1}^{x} \hat{S}_{3k}^{x} \rangle
= K_3/4 + m_0 L_3 + \varepsilon_0 M_3, \label{eq:qhx} \\
c_{\rm ih}^{zz} &\equiv&                    
\langle \hat{\mu}_{3k-2}^{z} \hat{S}_{3k-1}^{z} \rangle
= L_1/8 + m_0 (K_1 + M_1)/4 + \varepsilon_0 L_1/2 \label{eq:Cihz}, \\
q_{\rm hh}^{zz} &\equiv&                     
\langle (\hat{S}_{3k-1}^{z})^2 \rangle 
= K_4/4 + m_0 L_4 + \varepsilon_0 M_4.  \label{eq:qhzz}
\end{eqnarray}
As one can see, all aforelisted quantities can be expressed in terms of the single-site magnetization 
$m_0$ and the nearest-neighbour correlation $\varepsilon_0$ of the spin-$\frac12$ Ising linear chain 
given by $R$ and $H_0$. Since exact analytical formulae for those quantities have been retrieved in the literature many times before \cite{Lav99}, we shall restrict ourselves for brevity to listing the 
coefficients $K_i$, $L_i$ and $M_i$ ($i$ = 1--3) emerging in the set of Eqs. (\ref{eq:miz})-(\ref{eq:qhzz})
\begin{eqnarray} 
\label{eq:KLM}
K_i &=& F_i(\beta J_{\rm I}) + F_i(- \beta J_{\rm I}) + 2F_i(0), \label{eq:KLM1} \\
L_i &=& F_i(\beta J_{\rm I}) - F_i(- \beta J_{\rm I}),   \\
M_i &=& F_i(\beta J_{\rm I}) + F_i(- \beta J_{\rm I}) - 2F_i(0). \label{eq:KLM4}  
\end{eqnarray}
An explicit representation of the functions $F_i(x)$ is too cumbersome to write 
it down here and it is therefore left for Appendix.

Finally, we just simply quote the well-known thermodynamical-statistical relations, which have 
been utilized for calculating Gibbs free energy ${\cal G}$, entropy $S$, specific heat $C$ and 
susceptibility $\chi$. Accurate results for these quantities have been obtained with the help 
of the precise mapping relation (\ref{eq:ZdZ0}) and using
\begin{eqnarray}
{\cal G} = - k_{\rm B} T \ln {\cal Z} = {\cal{G}}_0 - N k_{\rm B} T \ln A, \\
S = -\Big( \frac{\partial {\cal G}}{\partial T} \Big)_{H}, \qquad
C = -T \Big(\frac{\partial^2 {\cal G}}{\partial T^2} \Big)_{H}, \qquad
\chi = - \Big(\frac{\partial^2 {\cal G}}{\partial H^2} \Big)_{T},
\end{eqnarray}
where ${\cal G}_0$ is referred to as the Gibbs free energy of corresponding spin-$\frac12$ Ising chain \cite{Lav99}.

\section{Results and Discussion}

Before proceeding to a discussion of the most interesting results it is worthy to mention that the results  derived in the preceding section hold regardless whether ferromagnetic or antiferromagnetic interactions 
are assumed. As we are mainly interested in examination of the spin frustration effect, in what follows 
we restrict both exchange parameters $J_{\rm H}$ and $J_{\rm I}$ to positive values in order to match the situation in the frustrated antiferromagnetic diamond chain. Furthermore, 
it is convenient to reduce the number of free parameters by assuming equal $g$-factors for the Ising 
and Heisenberg spins, i.e. by imposing the same parameter representing the effect of external field 
$H_{\rm I} = H_{\rm H} = H$. To simplify further discussion, we shall also introduce a set of reduced parameters $t = k_{\rm B}T/J_{\rm I}$, $h = H/J_{\rm I}$ and $\alpha = J_{\rm H}/J_{\rm I}$ 
as describing dimensionless temperature, external field and the strength of frustration, respectively. 

\subsection{Spin-$\frac12$ Ising-Heisenberg diamond chain}

First, let us take a closer look at the ground state of the spin-$\frac12$ Ising-Heisenberg diamond chain. 
For illustration, two ground-state phase diagrams are displayed in Fig. 2, the one in the $\Delta - \alpha$ space for the system without external magnetic field (Fig. 2a) and the other one in the $\alpha - h$ space 
for the system placed in the non-zero external field under the assumption $\Delta = 1.0$ (Fig. 2b). 
Both the figures clearly demonstrate that a competition between the interaction parameters $\alpha$, $\Delta$ and $h$ gives rise to three possible ground states. Apart from the usual ferrimagnetic phase (FRI) and the frustrated phase 
(FRU) found both in a presence as well as absence of the external field, the system ends up in the saturated paramagnetic phase (SPP) once the external field is being above its saturation value. Spin order within 
FRI, FRU and SPP can be distinguished from one another according to their attributes (physical quantities included in the set of Eqs. (\ref{eq:miz})-(\ref{eq:qhzz})), as well as, through their wave functions  
\begin{eqnarray}
\fl |FRI \rangle =  \prod_{k=1}^{N} | - \rangle_{3k-2} 
\prod_{k=1}^{N} | +, +  \rangle_{3k-1, \, 3k}, \nonumber \\
m_{\rm i}^z = -0.5, m_{\rm h}^z = 0.5, c_{\rm hh}^{zz} = 0.25, 
c_{\rm hh}^{xx} = 0, c_{\rm ih}^{zz} = -0.25; 
\label{FRIa}	
\end{eqnarray}
\begin{eqnarray}
\fl |FRU \rangle = \prod_{k=1}^{N} |\pm \rangle_{3k-2} \prod_{k=1}^{N} \frac{1}{\sqrt{2}} 
\Bigl( |+, - \rangle - |-, + \rangle \Bigr )_{3k-1, \, 3k},
\nonumber \\   
m_{\rm i}^z = 0, m_{\rm h}^z = 0, c_{\rm hh}^{zz} = -0.25, 
c_{\rm hh}^{xx} = -0.25, c_{\rm ih}^{zz} = 0; 
\label{FRUa}	
\end{eqnarray} 
\begin{eqnarray}
\fl |SPP \rangle =  \prod_{k=1}^{N} | + \rangle_{3k-2} 
\prod_{k=1}^{N} | +, +  \rangle_{3k-1, \, 3k}, \nonumber \\
m_{\rm i}^z = 0.5, m_{\rm h}^z = 0.5, q_{\rm hh}^{zz} = 0.25, c_{\rm hh}^{zz} = 0.25, 
c_{\rm hh}^{xx} = 0, c_{\rm ih}^{zz} = 0.25.
\label{SPPa}	
\end{eqnarray}
The first product in the aforelisted eigenfunctions is carried out over all Ising spins, the second one 
runs over all pairs of the Heisenberg dumbbell spins and $|\pm \rangle$ denotes standard ket vector assigned to $z$th projection of the Ising ($\mu^z = \pm \frac12$) and Heisenberg ($S^z = \pm \frac12$) spins. 
It is quite evident from the set of Eqs. (\ref{FRIa}) that FRI displays the classical ferrimagnetic spin arrangement usually observed in the pure Ising systems, actually, all the results clearly indicate antiparallel alignment between the nearest-neighbouring Ising and Heisenberg spins. It should be emphasized, however, that the classical ferrimagnetic order originating from the antiferromagnetic Ising interaction $J_{\rm I}$ can be destroyed through the competing Heisenberg interaction $J_{\rm H} (\Delta)$ that brings 
a frustration into play. As a matter of fact, the spin order dramatically changes when the frustration 
parameter $\alpha$ exceeds the boundary value  
\begin{eqnarray}
\langle FRI | FRU \rangle: \alpha_{\rm b} = \frac{2}{\Delta + 1}.
\end{eqnarray}
In such case, all Heisenberg spin pairs create singlet dimers and on account of this singlet pairing, 
all Ising spins become completely free to flip. In other words, the Ising spins are thoroughly uncorrelated 
in FRU on account of the frustration arising from the singlet pairing between the Heisenberg spins 
as also suggested by (\ref{FRUa}). 
Owing to this fact, it might be concluded that the diamond spin chain splits into a set of independent monomers (Ising spins) and dimers (Heisenberg spin pairs) whenever the frustration parameter is stronger 
than its boundary value $\alpha_{\rm b}$. Thus, FRU is virtually being macroscopically degenerate monomer-dimer state with the residual entropy $S_{\rm res}/3N = \ln(2)^{1/3}$ proportional to the 
total number of frustrated Ising spins. For completeness, it should be also noticed that 
sufficiently strong external field stabilizes the standard SPP regardless whether FRI or FRU constitutes 
the zero-field ground state (see Fig. 2b). As could be expected, all Ising as well as Heisenberg spins 
tend to align into the external-field direction above the saturation field, which represents lower bound 
for an occurrence of SPP.

Next, we turn our attention to the magnetization process at zero as well as non-zero temperatures. 
For this purpose, two typical magnetization vs. field dependences are plotted in Fig. 3 for several dimensionless temperatures. It can be readily understood by comparing the displayed magnetization 
curves with the phase diagram shown in Fig. 2b that two different zero-temperature limits obviously 
reflect both possible types of field-induced transitions FRI-SPP (Fig. 3a) and FRU-SPP (Fig. 3b). 
Since the ground state is being formed in the former (latter) case by FRI (FRU), the zero-temperature magnetization curve depicted in Fig. 3a (Fig. 3b) starts from non-zero (zero) magnetization in the 
limit of vanishing external field. Contrary to this, both magnetization curves always start from 
zero magnetization (disordered state) at any finite temperature according to the one-dimensional 
character of the investigated spin system. It should be emphasized, moreover, that the magnetization 
jumps to be observed in the magnetization curves strictly at $t = 0$ are gradually smeared out 
when temperature is raised from zero. In addition, the higher the temperature, the smaller the width 
of magnetization plateaux (horizontal regions in the magnetization vs. field dependence), which
entirely disappear from the magnetization curves above a certain temperature. Finally, it is quite 
interesting to mention that the identified magnetization plateaux at one third of the saturation 
magnetization satisfy the Oshikawa-Yamanaka-Affleck rule \cite{Osh97} proposed for the formation 
of quantized plateaux.

Now, let us step forward to a discussion concerned with the thermal dependence of the zero-field susceptibility times temperature ($\chi t$) data as displayed in Fig. 4. If the frustration parameter 
$\alpha$ is selected so that FRI constitutes the ground state (Fig. 4a), then, $\chi t$ data 
exhibit a round minimum prior to exponential divergence appearing on temperature decrease. 
As it can be clearly seen from Fig. 4a, the stronger the frustration parameter $\alpha$, 
the deeper the notable minimum whose position is simultaneously shifted towards lower temperatures. 
It should be mentioned, moreover, that an appearance of the round minimum in the $t-\chi t$ dependence 
is a typical feature of quantum ferrimagnets, since $\chi t$ product monotonously decreases with 
temperature for ferromagnets, while it monotonously increases for antiferromagnets \cite{Yam98}. 
Accordingly, a location of the round minimum can be regarded as a point that determines ferromagnetic-to-antiferromagnetic crossover in view of thermal exitations. On the other hand, 
if the frustration parameter $\alpha$ drives the system into the disordered FRU ground state (Fig. 4b), 
the $\chi t$ product then exhibits a round minimum before it tends to the constant value $1/12$ by
 approaching zero temperature. Notice that this zero-temperature value can be explained in compliance 
with the Curie law of the frustrated Ising spins, which effectively form isolated spins (monomers) in FRU. 

Another quantity, which is important for overall understanding of thermodynamics, is being the specific heat. Temperature variations of the zero-field specific heat are depicted in Figs. 5a-b for $\Delta = 1.0$ and several values of the frustration parameter $\alpha$. According to these plots, there still emerges at least one round maximum, which can be thought of as the usual Schottky-type maximum irrespective whether FRI 
or FRU constitutes the ground state. If the frustration parameter is selected sufficiently close to 
FRI-FRU phase boundary given by $\alpha_{\rm b}$, however, there also appears an additional second maximum located in the low-temperature part of the specific heat. Apparently, the low-temperature peak becomes the more pronounced, the closer is $\alpha$ selected to $\alpha_{\rm b}$. When the frustration parameter $\alpha$ drives the system into the disordered FRU ground state, then, the striking second maximum gradually disappears upon further strenghtening of $\alpha$, since the low-temperature peak shifts towards higher temperatures until it entirely merges with the high-temperature Schottky-type maximum (see the curves labeled as 
$\alpha = 1.1, 1.25$ and $1.5$). These observations would suggest that the double-peak structure in the specific heat curves originates from thermal excitations between the ground-state spin configuration and the ones close enough in energy to the ground state.    

The situation becomes even more intriguing on applying the external magnetic field. As one can see from 
Fig. 5c, the rising external field generally causes a gradual increase in the height of the low-temperature peak and moves it towards the higher temperatures. Similar trend is also seen in a change of the size 
and position of the high-temperature maximum, albeit this Schottky-like peak moves towards higher temperature more slowly than the low-temperature one. As a result, both maxima coalesce at a certain value of the external field and above this value, the specific heat exhibits just single rounded maximum (see for instance the curve $h = 0.5$). Apart from these rather trivial findings, a remarkable triple-peak specific heat curves can 
also be detected when small but non-zero external field is applied to the system driven by the frustration into the disordered FRU state (the case $h = 0.05$ in Fig. 5c). Besides two aforedescribed peaks, whose 
origin has been resolved earlier, there also appears an additional third peak to be located at lowest temperature. There are strong indications that an appearance of this additional sharp maximum can be 
explained through the field-induced splitting of the energy levels related to the frustrated Ising spins. 
In accordance with this statement, the insert of Fig. 5d clearly demonstrates how this peculiar third 
maximum gradually shifts towards higher temperatures with increasing the field strength until it 
coalesces with the second low-temperature peak (see the curve labeled $h = 0.1$).

At last, we shall briefly discuss the adiabatic demagnetization curves studied in connection with the 
enhanced magnetocaloric effect. Some interesting results for adiabatic processes keeping entropy constant 
are presented in Fig. 6 in the form of the temperature vs. external field dependence. Two depicted sets of demagnetization curves reflect two available adiabatic scenarios related to SPP$\rightarrow$FRI (Fig. 6a) 
and SPP$\rightarrow$FRU (Fig. 6b) transitions. Apparently, the maximal cooling rate emerges in the 
vicinity of critical fields and zero field, where zero temperature is in principle reached whenever 
the entropy is set equal to or less than its residual value $S_{\rm res}$. It should be pointed out 
that a relatively fast heating, which occurs when the external field is lowered from its critical value, prevents a practical use of the whole demagnetization curves for a cooling purpose. In addition, 
the cooling effect becomes of technological relevance only if a cooling rate exceeds the one 
of paramagnetic salts. From this point of view, the enhanced magnetocaloric effect is observed 
only if the frustration drives the system into disordered FRU ground state (Fig. 6b). Even under 
this condition, the cooling rate of paramagnetic materials is exceeded only if the entropy is chosen 
close enough to its residual value $S_{\rm res}$ and the external field is below $h \approx 0.05$.
This limitation would imply that temperatures in the sub-Kelvin range are in principle accessible 
provided that the frustrated diamond chain compound  with exchange constants of the order of few 
tents Kelvin, such as azurite, is used as refrigerant. 

\subsection{Mixed spin-$\frac12$ and spin-$1$ Ising-Heisenberg diamond chain}

In this part, we shall turn our attention to the mixed spin-$\frac12$ and spin-$1$ Ising-Heisenberg 
diamond chain with the aim to clarify the impact of integer-valued Heisenberg spins on the magnetic
behaviour of the frustrated diamond chain. We start our discussion repeatedly with the ground-state 
analysis. The phase diagram constructed in an absence of the external field (Fig. 7a) implies 
existence of three possible ground states. Besides the semi-classically ordered ferrimagnetic phase 
(FRI), there also appear the quantum ferrimagnetic phase (QFI) and the frustrated phase (FRU). 
FRI, QFI and FRU can be characterized by means of
\begin{eqnarray}
\fl |FRI \rangle =  \prod_{k=1}^{N} | - \rangle_{3k-2} \prod_{k=1}^{N} | 1, 1  \rangle_{3k-1, \, 3k}, \nonumber \\
m_{\rm i}^z = -0.5, m_{\rm h}^z = 1, q_{\rm hh}^{zz} = 1, c_{\rm hh}^{zz} = 1, c_{\rm hh}^{xx} = 0, 
c_{\rm ih}^{zz} = -0.5; 
\label{FRI}	
\end{eqnarray}
\begin{eqnarray}
\fl |QFI \rangle = \prod_{k=1}^{N} |- \rangle_{3k-2} 
   \prod_{k=1}^{N} \frac{1}{\sqrt{2}} \Bigl(|1,0 \rangle - |0,1 \rangle \Bigr)_{3k-1, \, 3k},
\nonumber \\   
m_{\rm i}^z = -0.5, m_{\rm h}^z = 0.5, q_{\rm hh}^{zz} = 0.5, c_{\rm hh}^{zz} = 0, 
c_{\rm hh}^{xx} = -0.5, c_{\rm ih}^{zz} = -0.25;  
\label{QFI}	
\end{eqnarray}  
\begin{eqnarray}
\fl |FRU \rangle = \prod_{k=1}^{N} |\pm \rangle_{3k-2} \prod_{k=1}^{N} \frac{1}{2 \sqrt{\delta}} \Bigl[ \sqrt{\delta + 1} \Bigl( |1,-1  \rangle + |-1,1  \rangle \Bigr) 
- \sqrt{2} \sqrt{\delta - 1} |0, 0 \rangle \Bigr ]_{3k-1, \, 3k},
\nonumber \\   
m_{\rm i}^z = 0, m_{\rm h}^z = 0, q_{\rm hh}^{zz} = -c_{\rm hh}^{zz} = (1 + \delta^{-1})/2, 
c_{\rm hh}^{xx} = -2 \Delta \delta^{-1}, c_{\rm ih}^{zz} = 0; 
\label{FRU}	
\end{eqnarray} 
where $\delta = \sqrt{1 + 8 \Delta^2}$, the first product in the aforelisted eigenfunctions is taken 
over all Ising spins ($|\pm \rangle$ stands for $\mu^z = \pm \frac12$) and the second product runs over all
pairs of the Heisenberg dumbbell spins ($|\pm 1, 0 \rangle$ is assigned to $S^z = \pm 1, 0$). 
Analytic expressions for the phase boundaries depicted in Fig. 6a read
\begin{eqnarray}
\langle QFI | FRI \rangle: \alpha_{\rm b1} = \frac{1}{\Delta + 1}; \quad \quad
\langle FRU | QFI \rangle: \alpha_{\rm b2} = \frac{2 \Delta - 1 + \delta}{2 \Delta (\Delta + 1)}.
\end{eqnarray}
The most significant difference between the two investigated diamond chains apparently 
rests in a presence of QFI located in between FRI and FRU.  This observation would suggest that 
the geometric frustration initially favours uprise of QFI before it finally energetically stabilizes FRU. Accordingly, it might be concluded that there may not occur in the mixed spin-$\frac12$ and spin-1 diamond chain a direct frustration-induced transition between FRI and FRU except the one observable in the Ising limit ($\Delta = 0$). 

Next, the ground-state phase diagram reflecting the effect of external field is shown in Fig. 7b. 
This phase diagram suggests that in response to the applied external field there also may arise   
the quantum ferromagnetic phase (QFO) and the saturated paramagnetic phase (SPP) besides the 
aforementioned FRI, QFI and FRU phases
\begin{eqnarray}
\fl |QFO \rangle = \prod_{k=1}^{N} |+ \rangle_{3k-2} 
   \prod_{k=1}^{N} \frac{1}{\sqrt{2}} \Bigl(|1,0 \rangle - |0,1 \rangle \Bigr)_{3k-1, \, 3k},
\nonumber \\   
m_{\rm i}^z = 0.5, m_{\rm h}^z = 0.5, q_{\rm hh}^{zz} = 0.5, c_{\rm hh}^{zz} = 0, 
c_{\rm hh}^{xx} = -0.5, c_{\rm ih}^{zz} = 0.25;  
\label{QFO}	
\end{eqnarray}  
\begin{eqnarray}
\fl |SPP \rangle =  \prod_{k=1}^{N} | + \rangle_{3k-2} \prod_{k=1}^{N} | 1, 1  \rangle_{3k-1, \, 3k}, \nonumber \\
m_{\rm i}^z = 0.5, m_{\rm h}^z = 1, q_{\rm hh}^{zz} = 1, c_{\rm hh}^{zz} = 1, c_{\rm hh}^{xx} = 0, 
c_{\rm ih}^{zz} = 0.5. 
\label{SPP}	
\end{eqnarray}
Before proceeding further, let us make few comments on all these possible ground states. 
Since FRI and SPP are commonly observed also in the semi-classical Ising spin systems,
we should therefore concentrate on QFI, QFO and FRU, in which quantum entanglement of the 
Heisenberg spin pairs indicates the quantum nature of these phases. It can easily be understood 
from Eqs. (\ref{QFI}) and (\ref{QFO}) that QFI and QFO are quite similar one to each other. 
As a matter of fact, all pairs of Heisenberg spins reside in both phases the same eigenstate 
and thus, the only difference between them consists in the spin arrangement of their Ising counterparts. 
In QFO, which is stable at stronger fields, the Ising spins tend to align towards the external field, 
whilst they are oriented in opposite to the external field in QFI, which is stable at relatively weaker 
fields. Notice that the quantum entanglement between the pairs of Heisenberg spins that occurs in QFI and QFO
can also be understood within the valence-bond-solid picture \cite{Aff87, Aff88}. If spin-$1$ sites 
are decomposed into two spin-$\frac12$ variables, then, one of the decomposed spins at each spin-$1$ site forms the singlet-dimer with its nearest-neighbouring spin-$1$ site, while the other one is polarized by 
the external field. As a result of this incomplete pairing, each spin-$1$ site effectively acts 
in QFI and QFO as it would be the spin-$\frac12$ variable. Further, it should be also remarked 
that all Ising spins are completely free to flip (frustrated) in FRU on behalf of a preferred antiferromagnetic alignment between each pair of the Heisenberg spins. Owing to this fact, FRU can 
be viewed as a state characterized by a complete randomization of the Ising spins, what consequently 
leads to a macroscopic degeneracy of FRU resembled in its residual entropy $S_{\rm res}/3N = \ln(2)^{1/3}$.

To clarify the magnetization scenario available for the mixed spin-$\frac12$ and spin-$1$ diamond chain, 
we depict in Fig. 8 all possible types of magnetization curves. In agreement with the phase diagram shown 
in Fig. 7b, there are in total four different types of magnetization curves reflecting the transitions 
FRI-SPP (a), QFI-FRI-SPP (b), FRU-FRI-SPP (c), and FRU-QFO-SPP (d). It should be stressed, nevertheless, 
that the system undergoes true transitions between those phases merely at zero temperature, where indeed emerge stepwise magnetization curves with abrupt change(s) of the magnetization at critical field(s). 
However, it is worthy to notice that there are no real magnetization jumps at any finite temperature 
and the sharp stepwise magnetization curves to be observed at zero temperature are gradually smeared out as temperature increases. It can be clearly seen from Fig. 8 that an increase in temperature actually shrinks 
also the width of plateaux until the plateau states completely disappear from the magnetization curves 
above a certain temperature. The most notable magnetization curves are those with the zero-field ground state corresponding to FRU as shown in Figs. 8c-d. According to these plots, 
it can easily be realized that all frustrated Ising spins tend to align to the external-field direction 
for any finite but non-zero external field provided that temperature is set to zero. At non-zero temperatures, on the contrary, the magnetization rises much more steadily in the vicinity of zero field in comparison with the magnetization curves having the long-range-ordered FRI (Fig. 8a) and QFI (Fig. 8b) phases in the ground state. Finally, it is also worthwhile to remark that the observed magnetization plateaux at $\frac15$ 
(QFI and FRU) and $\frac35$ (QFO) of the saturation magnetization satisfy the Oshikawa-Yamanaka-Affleck 
rule \cite{Osh97}.

Now, let us investigate in particular an influence of the spin frustration on thermal variations 
of the zero-field susceptibility times temperature ($\chi t$) data. The temperature dependence 
of $\chi t$ product is displayed in Fig. 9a for several values of the frustration parameter $\alpha$ 
that determine the ground state to be either of FRI or QFI type. Interestingly, $\chi t$ data then 
exhibit a round minimum upon cooling, which is followed by an exponentially steep increase that 
appears by approaching zero temperature. It is quite obvious from Fig. 9a that the stronger the 
parameter of frustration $\alpha$, the deeper the notable minimum whose position is simultaneously 
shifted towards lower temperatures. It is worthy to remember, moreover, that the minimum in the 
$t - \chi t$ plot is being a typical feature of quantum ferrimagnets \cite{Yam98} and also in our case, 
the minimum becomes especially marked by selecting $\alpha \in (\frac12, 1)$ when QFI constitutes 
the ground state. On the other hand, 
the temperature dependence of $\chi t$ data are depicted in Fig. 9b for the case when the frustration 
parameter $\alpha > 1$ drives the system into the disordered FRU ground state. Under these circumstances, 
the susceptibility diverges as $t^{-1}$ at low temperatures and $\chi t$ product tends to constant value 
1/12 nearby zero temperature. This value can be interpreted in terms of the Curie law of the frustrated 
Ising spins, since the spin arrangement that appears in FRU can be viewed as being composed of an independent 
set of the antiferromagnetic Heisenberg dimers and isolated Ising monomers, whereas the former ones do not contribute to $\chi t$ product in the zero-temperature limit.

To gain an insight into overall thermodynamics, let us turn our attention to a thermal behaviour 
of the specific heat. For this purpose, the zero-field specific heat is plotted against temperature 
in Figs. 10a-b for several values of the frustration parameter $\alpha$. It can be clearly seen 
from Fig. 10a that a rounded Schottky-type maximum observable at smaller values of $\alpha$ (e.g. 
$\alpha = 0.25$) gradually changes, as the frustration strengthens, to a striking dependence with the 
shoulder superimposed on this round maximum ($\alpha = 0.4$). It is noteworthy that the shoulder becomes 
the more pronounced, the closer is $\alpha$ selected to $\alpha_{\rm b 1} = 0.5$ determining the phase boundary between FRI and QFI. Under further increase of the frustration the shoulder merges with the Schottky-type maximum what eventually gives rise to a peculiar non-rounded maximum with almost 
constant value of the specific heat over the wide temperature range ($\alpha = 0.6$). Next, Fig. 10b 
shows how the specific heat recovers its double-peak structure when $\alpha$ approaches another phase 
boundary between QFI and FRU to appear at $\alpha_{\rm b 2} = 1.0$ (see for instance the curves for 
$\alpha = 0.75$ and $0.9$). In addition, it is quite evident from Fig. 10b that repeated strengthening 
of the frustration results in a suppression of the low-temperature peak until it finally merges with 
the high-temperature one. Note furthermore that the high-temperature peak has in general tendency 
to enhance in magnitude (both in height as well as in width) as $\alpha$ increases and its position 
is shifted towards higher temperatures. Altogether, it might be concluded that the remarkable double-peak structure of the specific heat arises just when the frustration parameter is close enough to a phase 
boundary. This result is taken to mean that the observed double-peak specific heat curves always 
originate from thermal excitations to a spin configuration rather close in energy to the ground state. 

Even more striking situation emerges by turning on the external field. The overall trends of the external 
field is to increase height and width of the low-temperature peak and to shift it towards higher 
temperatures until it coalesces with the higher-temperature maximum. It should be remarked, nevertheless, 
that the most notable dependences of the specific heat arise from when the frustration leads to 
the disordered FRU ground state. Besides the aforedescribed general trends illustrated in Fig. 10c, 
there also appears a remarkable kind of the specific heat curve with the triple-peak structure as 
particularly drawn in Fig. 10d and its insert. Apparently, the additional third peak observable at 
very low temperatures occurs on applying the small but non-zero external field. It is therefore quite
reasonable to conjecture that an origin of this low-temperature peak lies in the field-induced splitting 
of energy levels related to the frustration of the Ising spins. Actually, the stronger the external field, 
the greater the splitting caused by the external field and consequently, the position of this peak 
steadily shifts towards higher temperatures. Above certain external field, the additional third peak 
vanishes because of merging with the low-temperature peak observable also in the zero-field case. 

Finally, we shall close our discussion with an exploration of the adiabatic demagnetization examined
in connection with the enhanced magnetocaloric effect. Adiabatic processes keeping the entropy constant 
are plotted in Fig. 11 in the form of the temperature vs. external field dependence. Two displayed sets 
of demagnetization curves evidently reflect adiabatic change of temperature, which accompanies the 
transitions SPP$\to$FRI$\to$QFI (Fig. 11a) and SPP$\to$QFO$\to$FRU (Fig. 11b). When comparing these 
results with the ones formerly discussed for the spin-$\frac12$ diamond chain (Fig. 6), one easily 
finds some similarities between the two investigated diamond chains. Indeed, the most obvious drop 
in temperature is retrieved once again in the vicinity of critical fields and zero field, where zero temperature is in principle reached whenever the entropy is set equal to or less than its residual 
value $S_{\rm res}$. Moreover, the enhanced magnetocaloric effect with the cooling rate exceeding 
the one of paramagnetic salts occurs just as the disordered FRU phase constitutes the ground state
and the external field is below $h \approx 0.05$. By contrast, the most obvious difference between 
the two investigated spin systems consists in the greater diversity of the adiabatic process of 
the latter (mixed-spin) system, which exhibits dependences with two critical fields in addition to 
the ones with one critical field only.  

\section{Conclusion}

Exactly solvable frustrated spin systems are currently much sought after in the field of condensed 
matter physics, since they can serve as useful model systems for in-depth understanding of the 
effect of geometric frustration still not fully elucidated, yet. In the present work, we have provided 
rigorous results for one notable example of such system, the mixed spin-$\frac12$ and spin-$S$ Ising-Heisenberg diamond chain tractable within the generalized decoration-iteration map. It is worthy 
to mention that this rather simple model system has primarily been developed to predict and to 
understand the behaviour of insulating magnetic materials, in which the Heisenberg dimers interact 
with the Ising monomers in such a way that they form the diamond chain. Notice that the coordination 
polymers consisting of  the pairs of interacting transition-metal elements (Heisenberg dimers) coupled 
to the rare-earth elements (Ising monomers) represent perspective experimental realizations of 
the proposed system. Although we are not aware of any real coordination compound, which would meet 
this requirement, the recent progress in the design and controlled synthesis of the molecular-based 
magnetic materials supports our hope that it would be possible to prepare such polymeric chains 
in the near future. 

It is worthwhile to remark that a special emphasis is in our study laid on the investigation of 
geometric frustration generated by the competition between Heisenberg- and Ising-type interactions. 
Our results clearly demonstrate that an interplay between the frustration and quantum fluctuations, 
both arising from the Heisenberg exchange interaction, is being at the origin of interesting behaviour 
not commonly observed in the semi-classical Ising spin systems. Indeed, this interplay gives rise 
to several peculiar ground states with entangled states of Heisenberg spins and quantum effects 
turned out to play a substantial role in determining their finite-temperature properties, as well. 
Among other matters, we have found rigorous evidence for appearance of the quantized plateux in the 
magnetization curves, the round minimum in temperature dependence of the susceptibility times 
temperature data, the double-peak zero-field specific heat curves, the enhanced magnetocaloric effect
and so on. In our opinion, the most interesting finding to emerge from our study is a direct evidence 
of the triple-peak specific heat curve that appears when applying small external field to the system 
driven by the frustration into the disordered state. To the best of our knowledge, the discovery and 
possible explanation of the triple-peak structure in the specific heat curve has not been reported 
for any frustrated system hitherto.

Last but not least, it should be remarked that the relative ease of generalized mapping method used 
here implies a possibility of further extensions. Actually, this approach can straightforwardly be 
extended to account also for the single-ion anisotropy effect, the biquadratic exchange interaction, 
the next-nearest-neighbour interaction, the multispin interactions, etc. It is also noteworthy that 
the applied procedure is not constrained neither by the lattice topology and thus, it can be utilized 
for investigating the effect of geometric frustration on the planar Ising-Heisenberg lattices 
composed of the diamond-shaped units \cite{Str01, Str06}. In this direction continuous our next work.

\section{Appendix}

Before we list explicit expressions for the function $G(x)$ and $F_i(x)$ ($i$ = 1--4), which enter 
into the transformation formulae (\ref{eq:DIT})-(\ref{eq:AR}) and the set of Eqs. (\ref{eq:KLM1})-(\ref{eq:KLM4}), respectively, let us establish the notation $j_{\rm H} = \beta J_{\rm H}$, $h_{\rm H} = \beta H_{\rm H}$ and $\delta = \sqrt{1 + 8 \Delta^2}$. \\

(a) Expressions for the spin case $S = \frac12$:
\begin{eqnarray}
G(x) = 2 \exp(-j_{\rm H}/4) \cosh(x - h_{\rm H}) + 2 \exp(j_{\rm H}/4) \cosh(j_{\rm H} \Delta/2),
\nonumber 
\end{eqnarray}
\begin{eqnarray}
F_1(x) &=& - \exp(-j_{\rm H}/4) \sinh(x - h_{\rm H})/G(x),
\nonumber \\
F_2(x) &=& [\exp(j_{\rm H}/4) \cosh(j_{\rm H} \Delta/2) 
        - \exp(-j_{\rm H}/4) \cosh(x - h_{\rm H})]/\bigl[2G(x)\bigr], \nonumber \\
F_3(x) &=& - \exp(j_{\rm H}/4) \sinh (j_{\rm H}\Delta/2)/\bigl[2G(x)\bigr].
\nonumber 
\end{eqnarray}

(b) Expressions for the spin case $S = 1$:
\begin{eqnarray}
G(x) &=& 2 \exp(j_{\rm H}/2) \cosh(j_{\rm H} \delta/2) 
      + 2\exp(-j_{\rm H}) \cosh[2(x - h_{\rm H})] \nonumber \\    
     &+& \exp(j_{\rm H}) + 4 \cosh(x - h_{\rm H}) \cosh(j_{\rm H} \Delta),
\nonumber 
\end{eqnarray}
\begin{eqnarray}
F_1(x) &=& - \{2 \exp(-j_{\rm H}) \sinh[2(x-h_{\rm H})] + 2\sinh(x-h_{\rm H}) \cosh(j_{\rm H} \Delta)\}/G(x),
\nonumber \\
F_2(x) &=& \{2 \exp(-j_{\rm H}) \cosh[2(x-h_{\rm H})] - \exp(j_{\rm H}) 
        - \delta \mathcal{C} - \mathcal{S}\}/G(x),
\nonumber \\
F_3(x) &=& - \{2 \cosh(x - h_{\rm H}) \sinh(j_{\rm H} \Delta) + 4 \Delta \mathcal{S}\}/G(x),
\nonumber \\
F_4(x) &=& \{2 \exp(-j_{\rm H}) \cosh[2(x - h_{\rm H})] +  \exp(j_{\rm H}) \nonumber \\
       &+& 2 \cosh(x - h_{\rm H}) \cosh(j_{\rm H} \Delta) + \delta \mathcal{C} + \mathcal{S}\}/G(x),
\nonumber 
\end{eqnarray}
where $\mathcal{S} = \exp(j_{\rm H}/2) \sinh(j_{\rm H} \delta/2)/\delta$ and 
$\mathcal{C} = \exp(j_{\rm H}/2) \cosh(j_{\rm H} \delta/2)/\delta$. 

\ack
The authors express gratitude to scientific grant agencies VEGA and APVV
supporting this work under the grants VEGA 1/2009/05 and APVT 20-005204.

\newpage

\begin{large}
\textbf{Figure captions}
\end{large}

\begin{itemize}

\item [Fig. 1]
Diagrammatic representation of the mixed-spin Ising-Heisenberg diamond chain. 
The empty (filled) circles denote the lattice positions of the Ising (Heisenberg) spins,
the ellipse demarcates $k$th diamond-shaped plaquette. 

\item [Fig. 2]
(a) Ground-state phase diagram in the $\Delta-\alpha$ 
plane for the system in an absence of the external magnetic field; 
(b) Ground-state phase diagram in the $\alpha-h$ plane for $\Delta = 1.0$.

\item [Fig. 3]
The total magnetization reduced with respect to its saturation value versus the external magnetic field at various temperatures $t$ = 0.0, 0.1, 0.3, 0.5 in ascending order along the direction of arrows.

\item [Fig. 4]
Thermal dependence of the zero-field susceptibility times temperature data for 
$\Delta = 1.0$, $\alpha$ = 0.25, 0.5, 0.75, 0.9, 1.0 (Fig. 4a) and $\Delta = 1.0$, 
$\alpha$ = 1.1, 1.25, 1.5, 2.0 (Fig. 4b) in ascending order along the direction of arrows. 
For clarity, the case $\alpha_{\rm c}$ = 1.0 corresponding to the FRI-FRU phase boundary 
is depicted as broken line.

\item [Fig. 5]
Temperature variations of the specific heat when $\Delta = 1.0$ is fixed.
Figs. 5a-b illustrate the effect of frustration parameter $\alpha$ on the shape of zero-field 
specific heat and Figs. 5c-d display the effect of the applied external magnetic field when 
the frustration parameter $\alpha = 1.25$ drives the system into the disordered FRU state. 

\item [Fig. 6]
Adiabatic demagnetization in the form of temperature versus external field dependence for 
$\Delta = 1.0$, $\alpha = 0.5$ (Fig. 6a) and $\Delta = 1.0$, $\alpha = 1.5$ (Fig. 6b). 
For better orientation, broken curve depicts the dependence when entropy is fixed at 
the residual value $S_{\rm res}/3N = \ln(2)^{1/3}$ of FRU.

\item [Fig. 7]
(a) Ground-state phase diagram in an absence of the external field; 
(b) Ground-state phase diagram in the $\alpha-h$ plane for $\Delta = 1.0$.

\item [Fig. 8]
The total magnetization reduced with respect to its saturation value as a function 
of the external field at various temperatures $t$ = 0.0, 0.1, 0.3, 0.5 in ascending 
order along the direction of arrows.

\item [Fig. 9]
Thermal dependence of the zero-field susceptibility times temperature data for 
$\Delta = 1.0$, $\alpha$ = 0.25, 0.4, 0.5, 0.6, 0.75, 0.9, 1.0 (Fig. 9a) and $\Delta = 1.0$, 
$\alpha$ = 1.1, 1.25, 1.5, 1.75 (Fig. 9b) in ascending order along the direction of arrows. 
For clarity, the cases $\alpha_{\rm c1}$ = 0.5 and $\alpha_{\rm c2}$ = 1.0 corresponding 
to the phase boundaries between FRI-QFI and QFI-FRU, respectively, are depicted as broken lines.

\item [Fig. 10]
Temperature variations of the specific heat when $\Delta = 1.0$ is fixed.
Figs. 10a-b illustrate the effect of frustration parameter $\alpha$ on the shape of zero-field specific 
heat, whereas Figs. 10c-d display the effect of applied external field when the frustration 
parameter $\alpha = 1.25$ drives the system into the disordered FRU state. 

\item [Fig. 11]
Adiabatic demagnetization in the form of temperature versus external field dependence 
when $\Delta = 1.0$, $\alpha = 0.75$ (Fig. 11a) and $\Delta = 1.0$, $\alpha = 1.75$ (Fig. 11b). 
For clarity, broken curve depicts the dependence when entropy is fixed at the residual value 
$S_{\rm res}/3N = \ln(2)^{1/3}$ of FRU.

\end{itemize}

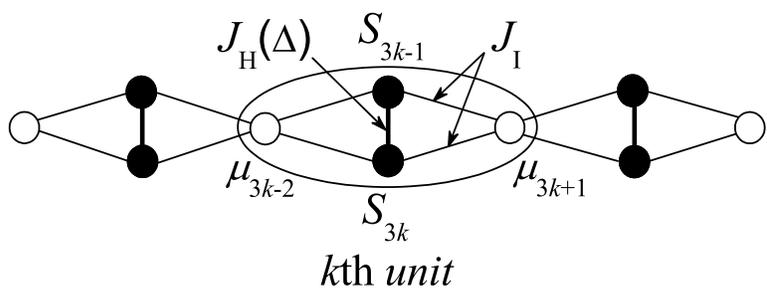
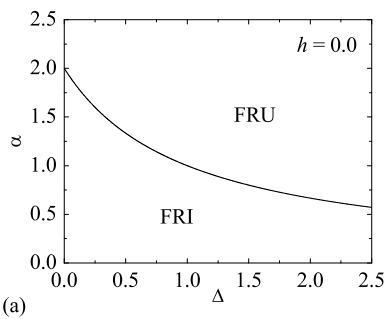
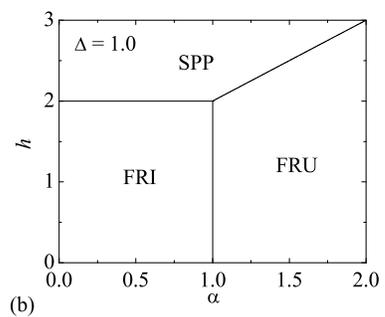
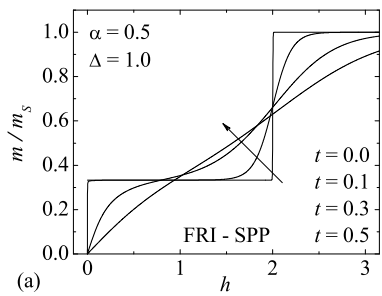
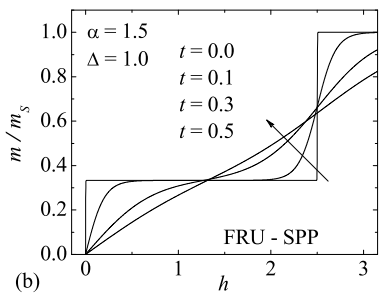
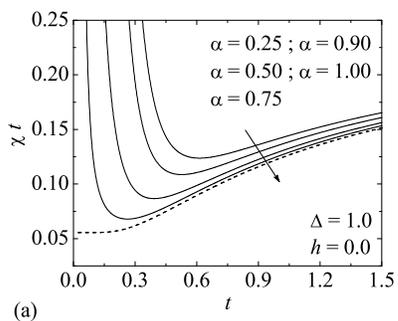
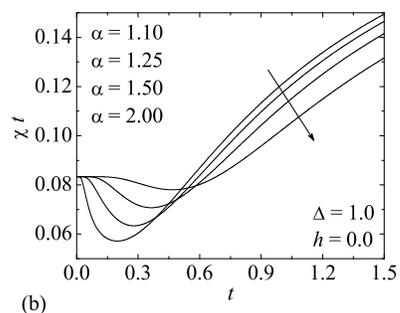
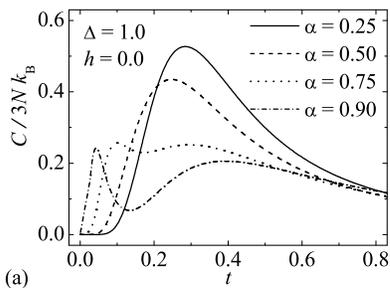
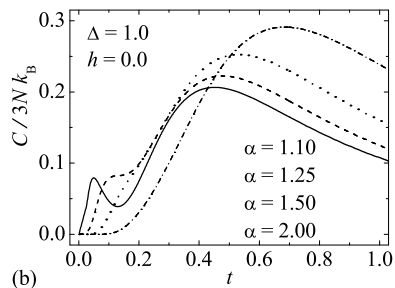
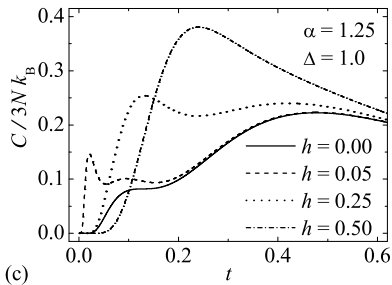
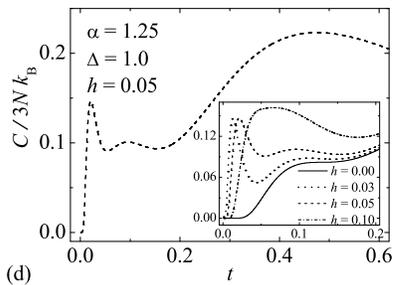
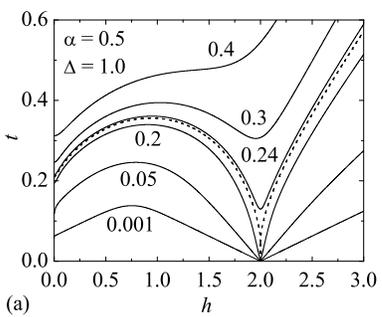
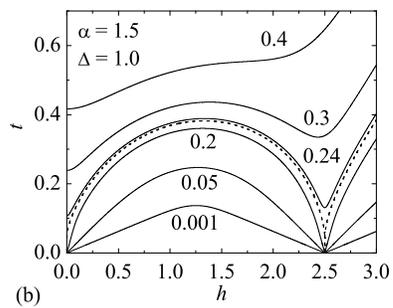
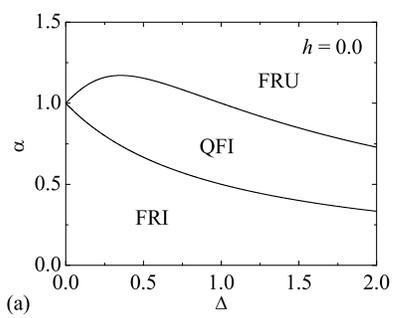
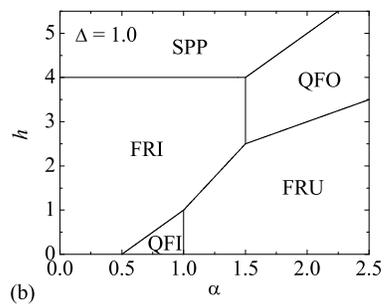
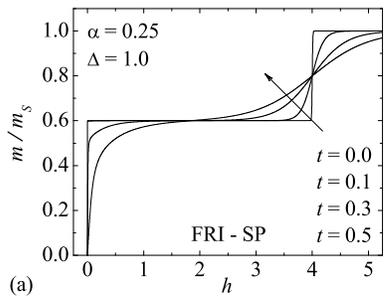
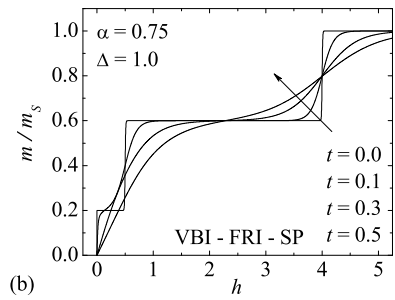
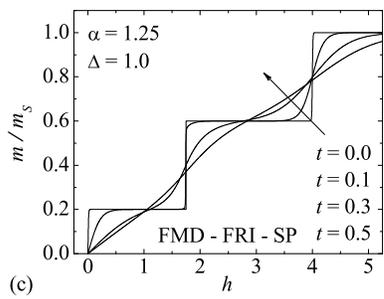
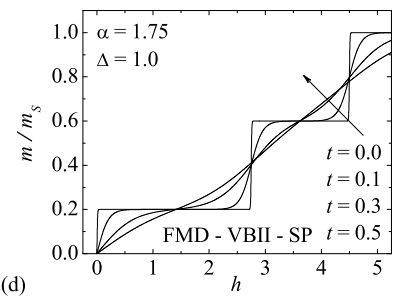
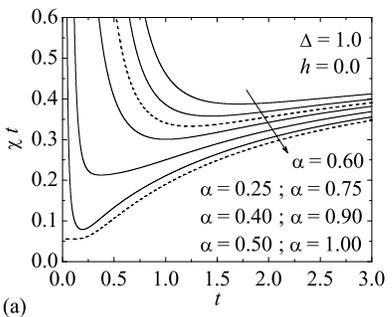
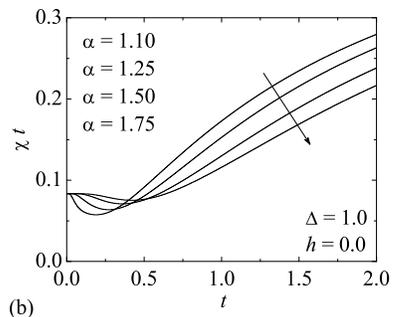
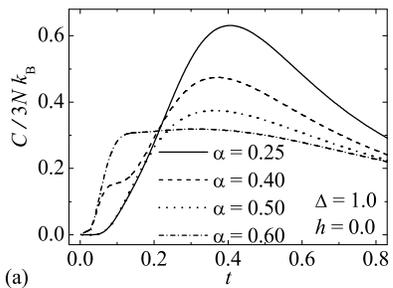
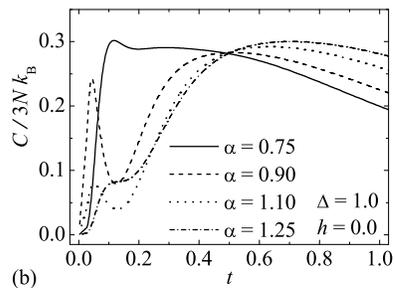
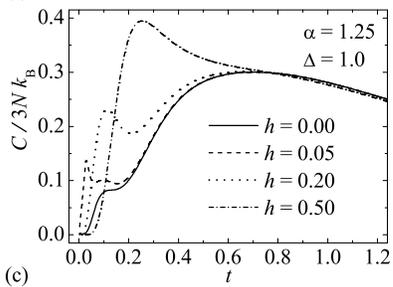
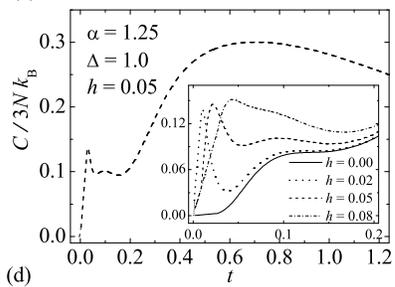
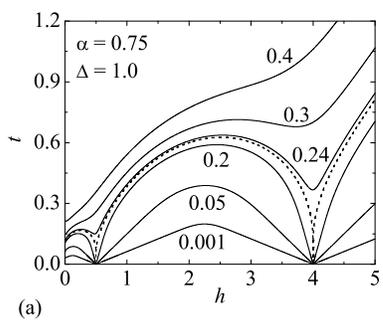
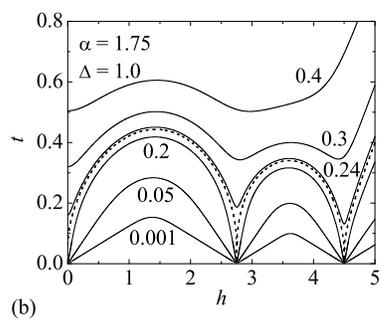
\end{document}